\documentclass[letterpaper]{article} 
\usepackage{aaai2026}  
\usepackage{times}  
\usepackage{helvet}  
\usepackage{courier}  
\usepackage[hyphens]{url}  
\usepackage{graphicx} 
\usepackage{amsmath}
\usepackage{natbib}  
\usepackage{caption} 
\usepackage{algorithm}
\usepackage{algorithmic}
\usepackage{comment}

\usepackage{newfloat}
\usepackage{listings}
\DeclareCaptionStyle{ruled}{labelfont=normalfont,labelsep=colon,strut=off} 
\floatstyle{ruled}
\newfloat{listing}{tb}{lst}{}
\floatname{listing}{Listing}
\nocopyright

\title{
Bio-inspired Learning and Decision-Making\\ with Probabilistic In-Memory Computing Hardware: Part 2
}

\author{
Thomas Dalgaty\textsuperscript{\rm 1},
Eiji Kawasaki\textsuperscript{\rm 1},
Miguel de Prado\textsuperscript{\rm 2, \rm 3},
Devendra Vyas\textsuperscript{\rm 2},
Tommaso Salvatori\textsuperscript{\rm 2 \rm 4},
Germain Haugou\textsuperscript{\rm 5},
Eric Flamand\textsuperscript{\rm 5}
}

\affiliations{
\textsuperscript{\rm 1}CEA-List, Grenoble, France
\textsuperscript{\rm 2} formerly VERSES AI Research Lab 
\textsuperscript{\rm 3} now PRAESC AI 
\textsuperscript{\rm 4} now TU Wien, Vienna, Austria
\\
\textsuperscript{\rm 5} Independent Researcher
\textsuperscript{\rm 6} Flacon, France

}

\begin{document}

\maketitle

\begin{abstract}
This report extends our previous work (Part 1), which introduced an energy-based model for learning and decision-making under uncertainty. The model leverages stochastic Langevin dynamics to continuously evolve approximate probability distributions over neuron states and model weights. However, as noted in Part 1 and confirmed through GPU-based implementations, large-scale probabilistic energy-based models of this nature face significant scalability challenges due to excessive execution latency.
This latency stems from a fundamental mismatch: massively parallel models with low arithmetic intensity (such as energy-based models) are being executed on processor architectures like GPUs that rely on high-bandwidth memory (HBM) interfaces. The HBM imposes brutally sequential execution constraints on inherently parallelizable models, creating the false impression that such models are unscalable. In reality, it is the GPU architecture itself, with its dependence on HBM interfaces, that is not a scalable processor architecture for this class of AI model.
In this report, we demonstrate using a detailed transaction-level model (TLM) of a probabilistic analogue in-memory computing (AIMC) processor that the same energy-based model can execute well over 1000× faster than data-center-grade hardware by eliminating the HBM interface and performing computation directly within on-chip memory.
\end{abstract}

\section{Introduction}

Graphics processing units (GPUs) have revolutionized the execution speed of algorithms with sequential computational patterns and high arithmetic intensity (i.e., those that reuse the same data loaded from memory multiple times), offering dramatic performance improvements over general-purpose CPUs. The GPU paradigm relies on storing massive quantites of model data off-chip in dynamic random access memory (DRAM) and transferring this data to on-chip static random access memory (SRAM) for processing. Typically, the SRAM capacity is approximately 1000× less than the DRAM \cite{nvidia2025blackwell}. This architecture has enabled the remarkable scalability of AI models such as convolutional neural networks (CNNs), previously considered to be un-scaleable, and has led to their profound and lasting impact on industry and society \cite{krizhevsky2017imagenet}.

A critical bottleneck in executing models like CNNs—often invisible in practice—is the interface between the external DRAM and the on-chip SRAM hierarchy on the GPU. In modern GPUs, this external memory and its interface are referred to as high-bandwidth memory (HBM) \cite{jun2017hbm}, named for the high-speed data transfer across thousands of lanes between the two circuits, achieving bandwidths on the order of terabytes per second (TB/s).

In response to the inherently low arithmetic intensity of large language model (LLM) inference, recent years have seen a surge of innovation in processor architecture and technology aimed at increasing model data bandwidth well beyond the TB/s regime. These advancements include: chiplet-level packaging that integrates multiple HBM and GPU chips on the same interposer; vertical integration of HBM chips on top of GPU chips \cite{vivet2020intact}; and architectural innovations such as near-memory computing \cite{caon2025scalable}, wafer-scale computing \cite{lie2024inside}, and in-memory computing \cite{hager202411,khaddam2022hermes}. The latter three approaches share a common philosophy: eliminating the HBM bottleneck by bringing model data as much as possible into on-chip memory, thereby enabling operator access with at least one order of magnitude higher bandwidth in a distributed fashion.

The ultimate solution in this direction is analogue in-memory computing (AIMC), where the same memory devices that store model data also perform computations—such as multiplication and addition—directly within the memory circuit itself, with extremely high bandwidth. However, AIMC introduces a significant challenge: storage and computation based on physical analogue phenomena are inherently subject to intrinsic and unavoidable noise and variability. These effects degrade data precision and significantly increase the bit-error rate during analogue in-memory computation. Recently, the field of probabilistic (or Bayesian) in-memory computing has gained considerable traction because it provides a means to exploit these intrinsic error sources to generate and compute with well-controlled probability distributions \cite{dalgaty2021situ,liu2022bayesian,lin2025deep,dalgaty2026memristors}. In this paradigm, noise and variability become integral parts of the algorithm, enabling reference software algorithms to run perfectly on analogue hardware.

In this technical note, we describe how the energy-based AI method introduced in Part 1 can be mapped and executed on a probabilistic AIMC processor. To achieve this, we implemented a full transaction-level model (TLM) of the AIMC processor and cross-compiled C++ code to run on the RISC-V cores within the TLM. Our results show that the energy-based model (EBM) described in Part 1 executes over 1000× faster on the AIMC processor model compared to GPUs and TPUs regularly found in datacentres.

\section{AIMC Processor Transaction-Level Model}
\label{sec:model}

We implemented the model using the open-source GVSOC framework \cite{bruschi2021gvsoc}, leveraging numerous open-source IP blocks from the PULP project.

At the lowest level of the AIMC processor's compute hierarchy are several 512×512 analogue memory arrays. In this report, we do not focus on a specific memory technology. By exploiting the intrinsic noise and variability of analogue memory devices, these cells, when written, generate and store in-place a sample encoded by a physical parameter of the cell. For example, the conductance of an oxygen-vacancy filament. Because the data is stored in a physical precision, we assume each sample can be represented as at least a 16-bit integer.

Four planes of these memory arrays share the same peripheral bank, which contains 128 digital-to-analogue converters (DACs) for writing to memory, 128 analogue-to-digital converters (ADCs) for reading from memory, and the drivers for selecting memory addresses. Adjacent to the four memory planes and the peripheral bank sits a RISC-V core. We use the open-source Spatz core \cite{cavalcante2022spatz} with a small L1 SRAM. An AXI bus connects the core to the analogue memory logic and a direct-memory access (DMA) controller, implemented using the open-source iDMA module from PULP \cite{benz2023high}, continuously cycles data between the analogue memory and the Spatz core at a rate that matches the extremely high bandwidth of the data produced/consumed by the analogue memory's ADCs/DACs.

We developed a pipeline that performs analogue writing and reading, DMA transfer, and RISC-V processing in parallel. Notably, the vector extensions of the Spatz core are essential for processing the data generated by the analogue memory's ADCs and preparing it for writing back to the analogue memory for multiplication or sampling operations via the DACs. Without these extensions, the RISC-V core would become a major bottleneck, negating the massive bandwidth available thanks to the analogue memory array.

At the next level of the hierarchy, multiple such cores are organised into a cluster. Another RISC-V core orchestrates the dispatch of compute jobs, the collection of results, and the reduction of results within the cluster, and includes its own L2 SRAM. These clusters are arranged in a 16×16 array and interconnected by a network-on-chip—we used the open-source FlooNoC from PULP \cite{fischer2025floonoc}. Amortised over a full processor model, comprising 10GB of analogue on-chip memory, the memory bandwidth of the analogue memory is in excess of some PB/s - between two and three orders of magnitude more than the HBM of the most recent GPUs.

The RISC-V cores adjacent to the analogue memory arrays write data via the DMA to a register file containing three sections: SAMPLE, DATA, and RESULT. The SAMPLE registers hold the data describing the probability distribution to be sampled (i.e., the mean and standard deviation of a Gaussian distribution). The DATA registers contain the input vector to be multiplied by the matrix of stored samples at the defined base address. The RESULT registers store the outputs of the vector-matrix operations. Once these registers are filled, the RISC-V core triggers operations in the analogue memory by writing to memory-mapped registers.

A low-level software driver library and a reactive software run-time model were implemented, enabling the development of C++ programs with calls to analogue functions that can then be cross-compiled using the riscv-toolchain to the TLM model to evaluate applications. While not cycle-accurate, the GVSOC framework enables rapid design exploration, provides access to many open-source IPs (such as those from the PULP project), and offers accurate latency estimations.

\section{EBM Model and Mapping Overview}
\label{sec:mapping}

The EBM is as described in Part 1 \cite{part1} and comprises five layers of 256 neurons each, replicated across 64 independent chains. Each chain contains 96 interacting particles, with each particle maintaining a sliding window of the most recent 512 samples. In total, the model requires 10 GB of memory.

In the context of active inference, for example, such a generative AI model would continuously update the probability distribution (approximated by 3 million samples) describing the agent's latent world model while simultaneously updating its internal generative model as it continuously observes, acts and learns from new data.

The EBM algorithm requires several key operations: evaluating scalar products to calculate Langevin gradients of internal neuron states; computing outer products for weight update gradients; performing element-wise operations to scale these quantities; and generating a massive number of Gaussian samples to add controlled noise to these gradients. By programming the analogue memory arrays with the 128 DACs in the peripheral bank, 128-element vectors can be sampled in parallel from specific distributions into designated addresses of the analogue memory array. These contents can then be read with a parallelism of 128 and processed by the Spatz core's vector extensions to perform element-wise scaling operations. The results can then be written to the register file as input data for operations such as scalar products with matrices of values already sampled inside the memory array.

We developed pipelined software kernels to implement the algorithm as described in Part 1. Both kernels follow this data flow pattern, constantly cycling outputs from the AIMC array through the Spatz core and back to the AIMC array, either to generate new samples or as input for vector-matrix products.

\section{Results and Outlook}
\label{sec:results}

We implemented C++ kernels for two primary operations: (i) generating one sample from the neuron distribution across all particles, and (ii) updating the weights of all particles based on the most recent sample from the neuron distribution in the sliding window.

The timing results for these two kernels are presented in Table \ref{tab:example}, with a comparison against the same kernels implemented on the Nvidia A100 GPU and Google v6e TPU as examples of data-center-grade processors based on HBM with similar memory capacities to our AIMC processor. For a fair comparison, both the GPU and TPU benchmarks use the 16-bit Bfloat data format, matching the 16 bits of data sampled in the analogue memory cells.

\begin{table}[h]
    \centering
    \caption{Execution time comparison for EBM kernels across different architectures.}
    \label{tab:example}
    \begin{tabular}{l c c c}
        \textbf{Kernel} & \textbf{Nvidia A100} & \textbf{Google TPU v6e} & \textbf{AIMC} \\
        Kernel (i) & 41 ms & 3186 ms & 21 $\mu$s \\
        Kernel (ii) & 14 ms & 3258 ms & 788 $\mu$s \\
    \end{tabular}
\end{table}

The execution latency of both kernels on the AIMC processor is at least three orders of magnitude faster than on both the GPU and TPU. Notably, even for the outer-product operation in kernel (ii)—where neither the AIMC processor nor the TPU's systolic array can exploit their full operator bandwidth—the AIMC processor remains vastly superior to the GPU. While impressive, this result simply reflects a fundamental architectural reality: processors relying on high-bandwidth interfaces to external DRAM cannot scale AI models with inherent massive algorithmic parallelism - such as EBMs and several other probabilistic models - whereas AIMC processors do so naturally. Furthermore, the AIMC's built-in ability to perform costly Gaussian probability distribution sampling and vector-matrix multiplication within the analogue memory using these samples, without further data movement, provides a structurally superior way to perform this computation that goes beyond the fundamental limits of digital memory technologies. 

\section{Acknowledgments}

This collaboration was supported by Horizon Europe's dAIEDGE network of excellence under Grant Agreement Number 101120726.

\bibliography{aaai2026,bib2}

\end{document}